# DFT study of rare earth (Tm, Yb, Ce) doped ZnO: structural, optoelectronic and electrical properties

M. Khuili[1,2,*], N. Fazouan[2,4], H. Abou El Makarim[3], E. H. Atmani[4], D. P. Rai[5] , M. Houmad[6]

[1] *Superior School of Technology (EST-Khenifra), University of Sultan Moulay Slimane, PB 170, Khenifra, 54000 Morocco*
[2] *Laboratory of Materials Physics, Faculty of Sciences and Technologies, B.P 523, 23000 Beni Mellal, Morocco.*
[3] *Laboratory LS3ME, Team Theoretical Chemistry and Molecular Modeling, University of Mohammed V Faculty of Sciences, Department of Chemistry, BP1014 Rabat, Morocco.*
[4] *Laboratory of condensed matters and renewables energies, Faculty of Sciences and Technologies, B.P  146, 20650 Mohammedia, Morocco*
[5]*Physical Sciences Research Center (PSRC), Department of Physics, Pachhunga University College, Aizawl-796001, India*
[6]*University Mohammed V Rabat Department of Physics-LMPHE*
[*]*E-mail:* khuilimohamed@gmail.com

**Abstract**

A comparative study of wurtzite ZnO doped by rare earth elements (Tm, Yb, Ce) have been investigated using density functional theory (DFT) based on the full-potential linearized augmented plane wave orbital (FP-LAPW) method, as implemented in Wien2K code. The structural parameters were calculated by PBEsol functional and in good agreement with the experimental data. The electronic (density of states, band structure) and optical (absorption coefficient, reflectivity, refraction index) properties were determined by TB-mBJ potential. The rare earth element doped ZnO have a significant impact on the optoelectronic properties which are mainly arise due to the presence of 4f electrons. The results of electronic structure shows that the doping of Tm, Yb, Ce on pristine ZnO has increases the band gap and in qualitative agreement with the experimental results. In many cases the Fermi level has been shifted to the conduction band, revealing n-type characters. Electrical conductivity has been calculated using BoltzTrap code based on the semi classical equation of Boltzmann. It has been observed that the conductivity has a direct relation with the temperature and carriers concentration. Our results provide the basis for future research in Tm, Yb, Ce doped ZnO compounds used as integrated optoelectronic devices and solar cells.

*Keywords*

ZnO, Rare Earth elements, DFT, Band gap, n-type characters, Optoelectronic devices,

Solar cells

# 1  Introduction

Zinc oxide (ZnO) is a transparent conductive oxide and it has been considered as a promising material for technological applications, such as photovoltaics, optoelectronics and sensing [1,2], in particular as a window for solar cells based on silicon as well as in lasers and light emitting diodes (LEDs) [3,4]. ZnO is characterized by its non-toxicity, its strong excitonic bond (60 meV) [5], its wide gap band of (3.3eV) and excellent transparency in the visible and near infrared ranges [6,7,8]. It is also characterized by its good thermal and chemical stability. Several theoretical and experimental studies have been perform on the doping of ZnO in order to obtained the enhanced electrical conductivity and broaden the spectrum of its transmittance. For this purpose, the group III elements like B, Al, Ga and In have been used extensively as an active members of dopant [9,10,11].

During the last twenty years, rare earths such as Yb, Tb, Eu, Ce, ..., have also been introduced into ZnO in order to obtain light emissions and absorptions at their proper wavelengths in the spectral ranges from UV to infrared. These dopings have various applications such as the conversion of photon energy in the photovoltaic field and, by combining the conductive nature of ZnO, for obtaining white light-emitting diodes [12,13,14]. Theoretically, it is well known that the 4f orbital of rare earth atoms also has an interesting peculiarity: its spatial extension is less important than the two 5s and 5p orbitals which are energetically inferior to it. This confinement of the 4f orbital is all the more important as we advance in the lanthanide series. Moreover, the electrons of 4f are not located at the theoretical position given by Klechkowski's rule. Indeed, the electrons of the 4f orbital should be placed at the periphery. However, in the lanthanide group, the 4f electrons are closer to the nucleus than those in the 5s and 5p orbitals due to a compression effect of the orbitals [15]. Thus, the shielding of the 4f orbital by the 5s and 5p orbitals makes it possible to limit the influence of an external crystalline field thus giving an emission stabilized according to the structure of the matrix. This phenomenon is not present for the transition metals which explain why for a single metal ion, we can obtain, unlike a rare earth ion, all the color variations according to the host matrix.

Recently, many works have been done on the doping of ZnO by rare earth elements. R. Vettumperumal et al. [16] prepared Er-doped ZnO nanostructures using the sol-gel technique, they revealed that ZnO doped by low-concentration of rare earth elements does not destroy the hexagonal structure of ZnO and exhibit a stronger excitonic bond than that of pure ZnO with a large transmission coefficient. Improving humidity detection performance based on Eu doped ZnO sensors has been studied by S. Yu et al. [17], by exposing them to humid environments in a wide range of 11 to 95% RH at room temperature. Their results show that this doping is a practical method for obtaining high humidity detection performance, which makes it a promising candidate for humidity detection materials. Thin films of ZnO doped with other rare earths such as Tm, Yb, Ce have been synthesized, for example ZZ. Gui et al. [18] using sol-gel method demonstrated that efficient organic solar cells can be obtained using zinc oxide doped with ytterbium ions as a cathode transport layer. H. Çola et al. [19] studied the synthesis and properties of undoped and Tm-doped ZnO nanowires obtained by the sol-gel and hydrothermal methods to determine its suitability for photovoltaics. They found that ZnO-NR was oriented on the c-axis and showed a significantly high optical transparency. In addition, they observed that electrical conductivity increases slightly with the quantity of doping; they concluded that Tm-doped ZnO-NR films can be used as transparent and conductive electrodes suitable for PV solar cells.

In this work, we did a comparative study of structural, optical and electrical properties of ZnO doped by three rare earth elements Ytterbium (Yb), Thulium (Tm) and Cerium (Ce). Our study is based on the Density Functional Theory (DFT) using the Linearized Augmented Plane Wave method in a Full Potential Linearized Augmented Plane Wave FP-LAPW. For the study of electronic and optical properties, we used the TB-mbj potential recently developed by Tran and Blaha [20] and to study the influence of dopants on transport phenomena such as electrical conductivity, we used Boltzmann's semi-classical equation. Our results are discussed and also compared with other experimental work.

## 2  Computational Details

In this paper, we studied the effect of doping ZnO by rare earth elements such as (Tm, Yb, Ce) using the FP-LAPW method implemented in the Wien2k code [21]. The idea of the method is that, far from the nuclei, the electrons can be considered as if they were free, and can be described by plane waves. While the electrons near the nucleus are closely bind by the electrostatic force and behave as if they were in isolated atoms. Thus the space in the unit cell can be separated into two regions (Figure 1):

- Muffin tin Spheres (MT) with Muffin tin Radius ($R_{MT}$) concentrated around the nuclei

and do not overlap. The wave function in this space is described by a linear combination of spherical harmonics.

- An interstitial region between the spheres. The wave function in this space is a plane wave function.

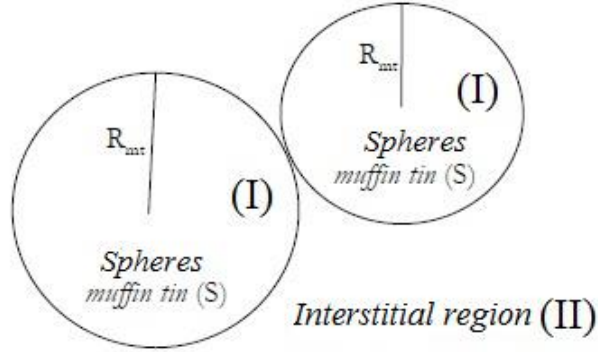

*Figure 1: Muffin-Tin (MT) sphere region and Interstitial region in the FP-LAPW method.*

So the wave function is of the form:

$$\varphi(r) = \begin{cases} \frac{1}{\sqrt{\Omega}} \sum_G C_G e^{i(G+K)r} & r \in (II) \\ \sum_{lm} A_{lm} U_l(r) Y_{lm}(r) & r \in (I) \end{cases} \quad (1)$$

Where $R_{MT}$ is the radius of the sphere MT, $\Omega$ the volume of the cell, $A_{lm}$ the coefficients of development in spherical harmonics $Y_{lm}$, and the function $U_l(r)$ is a regular solution of the Schrödinger equation for the radial part, $C_G$ a coefficient of the development in plane waves. The convergence of this basic set is controlled by the cutoff parameter $R_{MT}K_{max}$, where $R_{MT}$ is the smallest spherical atomic radius in the unit cell and $K_{max}$ is the magnitude of the largest wave vector $\vec{k}$ in the first irreducible Brillouin zone, $\vec{G}$ is the vector of the reciprocal space and $\vec{r}$ is the position inside the spheres.

In our calculations, we have chosen, $R_{MT}^{Zn}$=1.93au, $R_{MT}^{O}$=1.66au, $R_{MT}^{Ce}$=1.93au, $R_{MT}^{Tm}$=1.97au and $R_{MT}^{Yb}$=2.15au. The cutoff parameter of the plane wave in the interstitial region is chosen such that $K_{max} \times R_{MT} = 8$. The magnitude of the largest vector G in the Fourier expansion is $G_{max} = 16$ Ry. The energy convergence criterion is taken equal to 0.0001Ry. The revised functional PBEsol [22] given by Perdew-Burke-Ernzerhof is adopted to treat the exchange interaction between electron-electron during the structural optimization of all compounds. The supercell adopted in our calculations is of size 2x2x2 in which we substituted a Zinc atom by a rare earth atom to obtain the three structures of ZnO doped Yb, Tm and Ce at a rate of 6.25%. The electronic structure, the absorption coefficient and the refractive index were calculated using the TB-mbJ potential. The electrical properties such as the electrical conductivity and the density of the charge carriers of the doped structures were obtained using the BolzTrap code [23] which is based on the semi-classical Boltzmann equation combined with the electronic structure calculated from the code Wien2k [24].

## 3   Results and discussions

### 3.1   Structural parameters

ZnO crystallizes in the wurtzite structure which is considered to be the energetically most favourable as, compared to the zinc blende and Rocksalt structures [25]. Its unit cell contains two zinc atoms situated at (1/3,2/3,0), (2/3,1/3,1/2) and two oxygen atoms situated at (1/3,2/3,u), (2/3,1/3,1/2+u), where u is the internal parameter defined by the length of the Zn-O bond along the axis (0001) carried by c . The rare earth doped ZnO structures were generated by multiplying the primitive ZnO cell in three directions and replacing a zinc atom with a rare earth atom as illustrated in Figure 2. Before calculating the optical and electrical properties of doped ZnO structures, the structures were optimized by force minimization to obtain the new relaxed coordinates.  The structural optimization after the insertion of impurities in ZnO is a crucial step. For this effect, we start by relaxing each structure in the three directions X, Y and Z and we have calculated the energy of the structure according to its volume. This energy is then fitted by the Birch-Murnaghan equation of states [26] from which we deduced the optimum structural parameters as well as the compression modulus and its derivative.

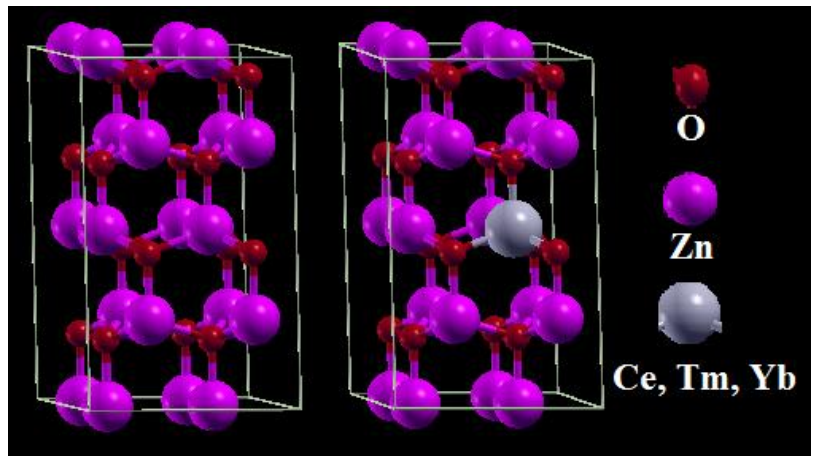

*Figure 2: 2x2x2 supercell of pure ZnO and ZnO doped with rare earth at 6.25%*

In Table 1, we have listed the structural parameters of equilibrium of doped ZnO structures deduced from our calculations and compared with other experimental work. From the results, it is noted that the parameters *a* and *c* of the doped structures increases as compared to those of undoped ZnO. This increase in structural parameters can be attributed to the difference between the ionic radius of $Zn^{2+}$(0.74 Å) and the ionic radii of the doping elements of $Ce^{3+}$(1.03 Å), $Tm^{3+}$(0.87 Å), and $Yb^{3+}$(0.86 Å). The three dopants have a larger ionic radius as compared to that of $Zn^{2+}$. This result is confirmed experimentally by [27,28,29] as illustrated in Table 1.

*Table 1: theoretical and experimental structural parameters of the rare earth doped ZnO*

| compound | Doping concentration | Method | a(Å) | c(Å) | Reference |
|---|---|---|---|---|---|
| **Pure ZnO** | 0% | This work | 3.232 | 5.225 | |
| | | Experimental | 3.24704 | 5.19912 | Lehlohonolo F. Koaoet Al [27] |
| | | Experimental | 3.25 | 5.20 | I. Soumahoro et Al [28] |
| | | Experimental | 3.2501 | 5.2069 | Hakan COLAK [29] |
| **Tm** | 6.25% | This work | 3.2787 | 5.285 | |
| | 6% | Experimental | 3.2645 | 5.2233 | Hakan COLAK |
| **Ce** | 6.25% | This work | 3.3089 | 5.322 | |
| | 10% | Experimental | 3.25052 | 5.20692 | Lehlohonolo F. Koao et Al |
| **Yb** | 6.25% | This work | 3.27555 | 5.283 | |
| | 5% | Experimental | 3.279 | 5.248 | I. Soumahoro et Al |

After optimizing the geometry of pure ZnO and rare earth doped ZnO, we simulated the X-ray diffraction spectra (XRD) using the code VESTA [30] to study the growth directions of these compounds. According to the results presented in Figure 3, the hexagonal structure of ZnO is still preserved even after the incorporation of the rare earth elements with no introduction of any secondary peaks or new phases. The XRD spectra show that the positions of the three main peaks (100), (002) and (101) of ZnO doped with rare earths move towards the small angles (see inset image in Fig. 3). This shift of peaks confirm the increase in the lattice parameters *a* and *c* inducing an extension of the structures of ZnO doped with rare earths.

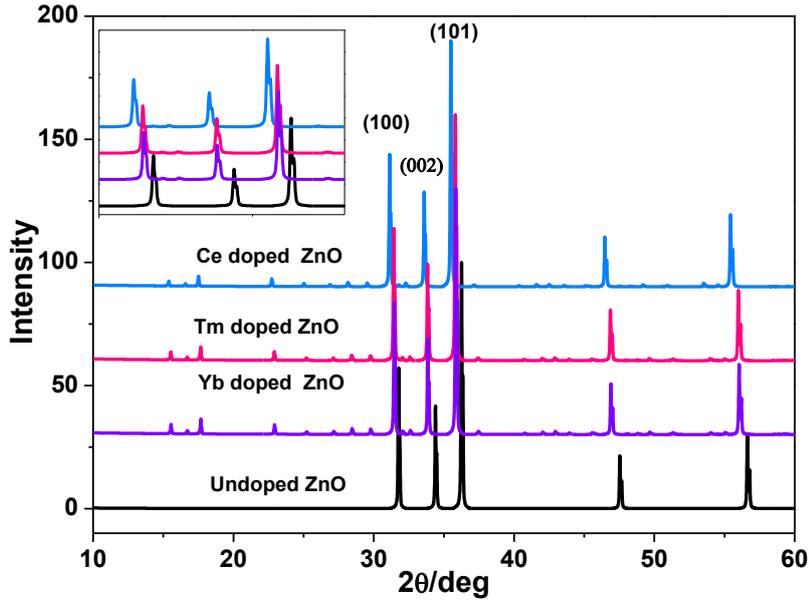

*Figure 3: DRX of pure ZnO and ZnO doped Tm, Yb and Ce at 6.25%*

## 3.2 Electronic structure

To study the electronic structure of rare earth doped ZnO, we have calculated the band structure, the partial and total density of states for all the compounds. We have used the functional GGA-PBEsol to approach the electron-electron exchange-correlation energy within the in the DFT approach. With this functional, the calculated gap energy is found to be 0.8 eV, which is far from the experimental gap. To remedy this underestimation of the gap, we introduced the TB-mbJ potential. The results of the structures of pure and doped ZnO bands are illustrated in Fig. 4. These figures show that the structures have a direct band gap whose value varies according to the type of doping. For pure ZnO, the direct energy gap determined at point Γ is 2.79 eV close to the experimental value of 3.3 eV. In the case of ZnO doped with rare earths, we have found that the Fermi level is shifted towards the conduction band. The presence of Fermi level close to the conduction band reflected that the system has now become n-type semiconductor. We have also noted the displacement of valence states towards the low energies with doping indicating the interaction between the atoms. The gap in doped system is measured between the Fermi level which is referred to as 0 eV and the maximum of the valence band [31]. From the band structure of each compound, we can clearly see the increase in the gap of 2.79 eV in the case of pure ZnO, 3.33, 3.35 and 3.79 eV for the ZnO doped with Yb, Tm and Ce respectively. This increase in the gap is attributed to the Burstein–Mott effect [32,33]. The appearance of new occupied states between the conduction band and the Fermi level attributed to the 4f orbitals of rare earth dopants. Thus, all the photons having a low energy can be absorbed by the doped ZnO compound, relatively increasing the absorption coefficient in the range of visible light.

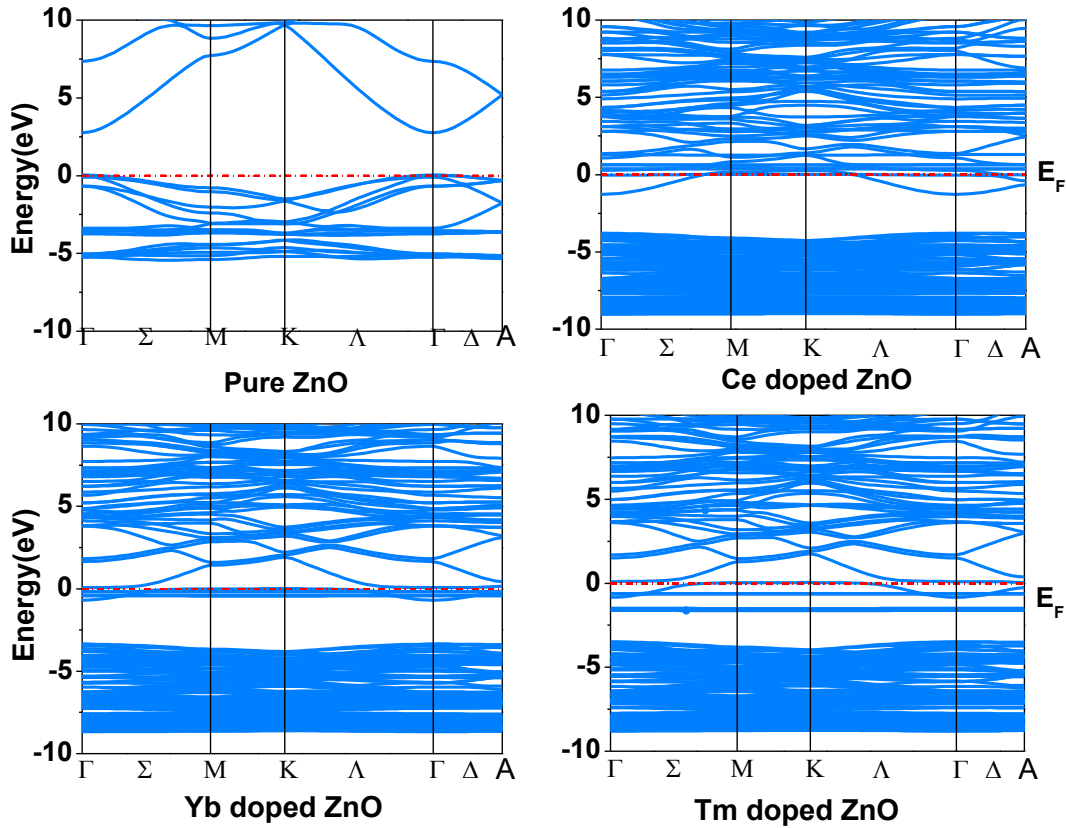

*Figure.4. the band structures of pure ZnO and ZnO doped Ce, Yb and Tm*

To clarify the origin of these band structures, we have calculated the partial density (PDOS) and total density of states (DOS) of pure and doped ZnO and presented in Figure 5. For pure ZnO, the valence band can be divided into two regions, the lower region of the band located around -18eV, consists of the O-2s states and the upper region between - 5.7 and 0 eV shows the contribution of the Zn-3d and O-2p orbitals. The conduction band is mainly made up of O-2p and Zn-4s orbital. However, the doping of ZnO by the rare earth elements has shifted the valence band towards low energies as compared to pure ZnO. In addition the presence of occupied states around the Fermi level which will essentially determine the properties optoelectronics of these structures.

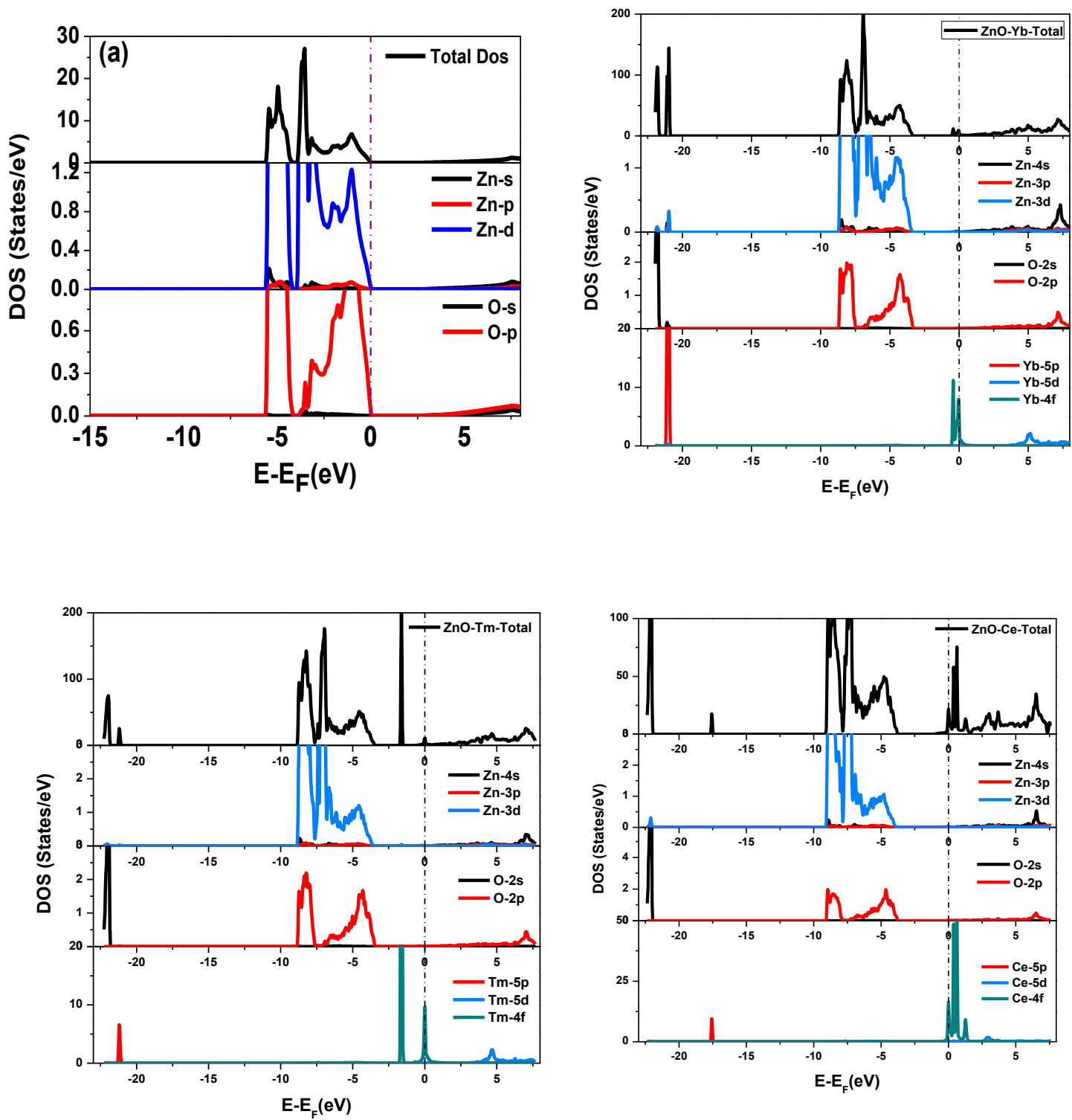

*Figure.5. the partial density (PDOS) and total density (DOS) of the states of pure and doped rare earth ZnO*

## 4  Optical properties

In the Wien2k code, all the optical properties such as the absorption coefficient, the transmittance and the refractive index of each doped ZnO structure are determined from the dielectric function ε. These optical properties are expressed as a function of frequency or wavelength. The dielectric function consists of two parts $\varepsilon_1$ and $\varepsilon_2$.

$$\varepsilon(\omega) = \varepsilon_1(\omega) + i\varepsilon_2(\omega) \qquad (2)$$

$\varepsilon_1$ and $\varepsilon_2$ which are given by the following equations [34] [35]:

$$\varepsilon_2(\omega) = \frac{8}{2\pi\omega^2}\sum_{nm'}\int |P_{nn'}(k)|^2 \frac{dS_k}{\nabla\omega_{nn'(k)}} \qquad (3)$$

$$\varepsilon_1(\omega) = 1 + \frac{2}{\pi}P\int_0^\infty \frac{\omega'\varepsilon_2(\omega')}{\omega'^2 - \omega^2}d\omega' \qquad (4)$$

$P_{nn'}(k)$ is the dipole matrix element, $\omega_{nn'(k)}$ is the energy difference between the initial and final states, $S_k$ represent the constant of a surface energy and p is the main part of the integral. The absorption coefficient α is proportional to the imaginary part of $\varepsilon_2$:

$$\alpha(\omega) = \frac{\omega\varepsilon_2(\omega)}{c} \qquad (4)$$

The refractive index is a function of $\varepsilon_1$ and $\varepsilon_2$

$$n(\omega) = \frac{1}{\sqrt{2}}\left[\left(\varepsilon_1^2(\omega) + \varepsilon_2^2(\omega)\right)^{\frac{1}{2}} + \varepsilon_1(\omega)\right]^{1/2} \qquad (5)$$

The reflectivity is calculated by the following relation [36]:

$$R(\omega) = \left|\frac{\tilde{n}-1}{\tilde{n}+1}\right| = \frac{(n-1)^2 + k^2}{(n+1)^2 + k^2} \qquad (6)$$

where k is the extinction factor.

Transparent conductive oxides (TCO) are known in the visible range by their low absorption and reflectivity, their low refractive index and their high rate of transmittance. Figure 6 shows the absorption coefficient for pure ZnO and rare earth doped ZnO. According to the spectra, it can be seen that pure ZnO has a low absorption in the visible and near infrared domains, with an increase in the ultraviolet range. After the doping of ZnO by the rare earth elements, ZnO exhibit relatively low absorption coefficient due to the creation of new states between the Fermi level and the valence band allowing the absorption of low energy photons. A blue shift in the absorption coefficient of doped system is observed and is related to the Burstein–Mott effect. This shift is well pronounced for doping in Ce as compared to that in Tm and Yb. This result is also observed experimentally by L. Arun Jose and Al who studied lanthanum-doped ZnO, of the same nature as the dopants that we used [37]. The reflectivity spectra of pure and doped ZnO show that all the structures have a low reflectivity which is less than 7% in the visible range. Further decrease of reflectivity spectra for doped structures has been observed. The Ce doped ZnO shows the lowest reflectivity as compared to all the dopants. In the near infrared and ultraviolet range, the reflectivity of ZnO doped with rare earths is reversed and becomes more pronounced than that of pure ZnO. The reduction in absorption and reflectivity in the visible domain of ZnO doped with rare earths resulted in broad transmittance allowing them to be good candidates for applications in the photovoltaic and also optoelectronic fields. The refractive index for all doped and pure ZnO has been calculated and presented in Fig. 6. It is clearly shown that the doping ZnO by the rare earths induces a reduction in this index signifying a low dispersion of light in these structures.

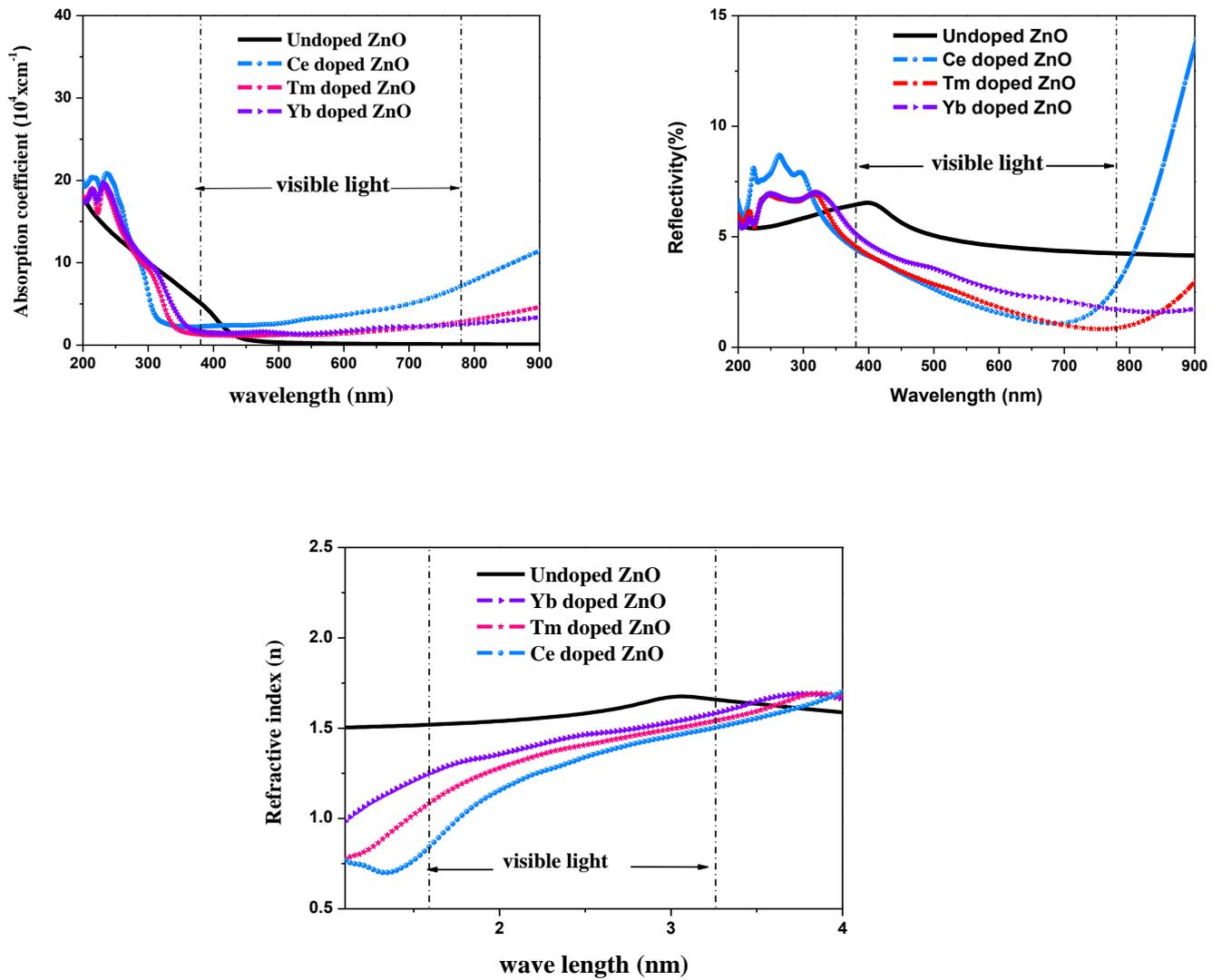

*Figure 6: Absorption coefficient, reflectivity and refractive index of pure and doped ZnO*

# 5 Electrical properties

Pure ZnO is characterized by its low electrical conductivity, for this it is preferable to be doped with elements which will improve its electrical conductivity and reduces its resistivity. In this part we have discussed the effect of inserting the elements like Ce, Tm and Yb in the ZnO. To fulfill this objective we have used the BoltzTrap code which is based on the semi classical Boltzmann equation combined with the results of the electronic structure from Wien2k. The code also makes it possible to study all the thermoelectric properties of materials as a function of temperature (T) and chemical potential (μ). The electrical conductivity is given by:

$$\sigma_{\alpha\beta}(T,\mu) = \frac{1}{\Omega} \int \sigma_{\alpha\beta}(\varepsilon) \left[-\frac{\partial f_\mu(T,\varepsilon)}{\partial \varepsilon}\right] d\varepsilon \qquad (7)$$

where $f_\mu$ is the fermi distribution, $\Omega$=volume of the supercelle and $\sigma_{\alpha\beta}(i,k)$ is the conductivity tensor:

$$\sigma_{\alpha\beta}(i,k) = e^2 \tau_{i,k} v_\alpha(i,k) v_\beta(i,k) \qquad (8)$$

$\tau$ is the electronic relaxation time which designates the time between two successive collisions of the charge carriers in the material, $v_\alpha$ and $v_\beta$ are the speeds of the charge carriers in the directions α and β. Fig. 7 shows the variation of electronic relaxation time dependent electrical conductivity (σ/τ) for different structures as a function of the temperature. From this figure, we can see the low value of electrical conductivity for pure ZnO has been improved after the insertion of the rare earth elements. This electrical conductivity grows rapidly as a function of temperature for all systems. However, a saturation of this conductivity for Ce doped ZnO is observed beyond 750K. It is also noted that Ce doped ZnO has a high electrical conductivity as compared to that of other structures. This behavior can be justified by the 4f states of Ce in the conduction band as compared to the same 4f states of Tm and Yb.

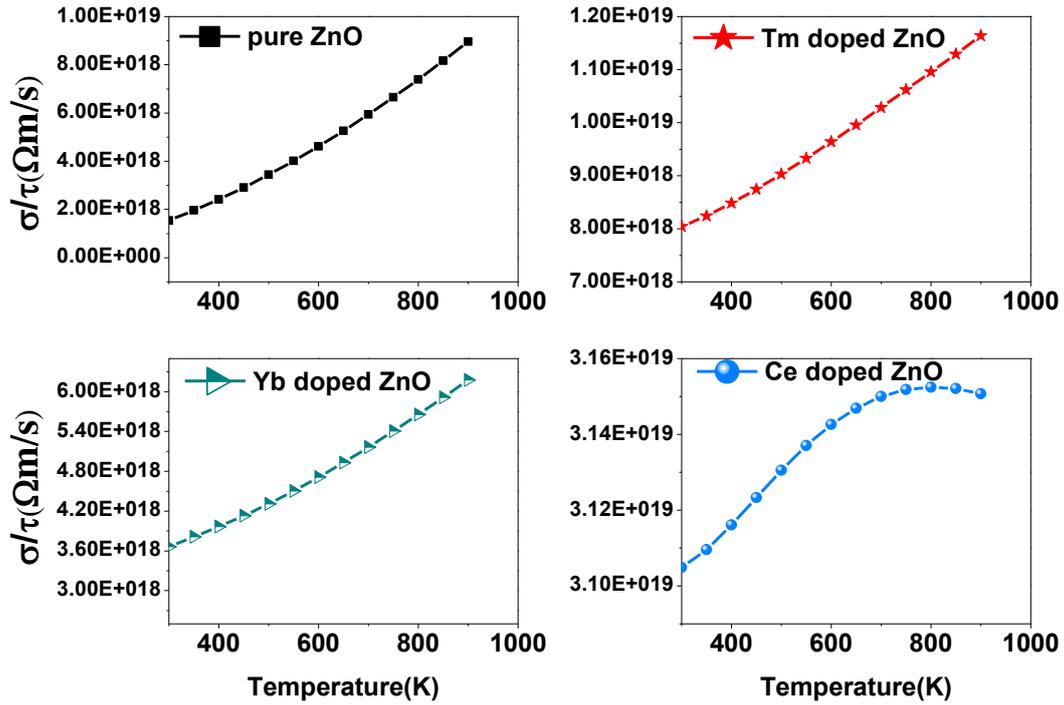

*Figure 7: electrical conductivity as function of* electronic *relaxation of time (σ/τ) as a function of temperature*

## 6 Conclusion

In this work we presented the results of doping of ZnO by the rare earth elements (Ce, Tm, Yb), using the DFT method. The results showed that the doping of ZnO by these elements at low concentrations does not distort the hexagonal symmetry of ZnO. However, we have reported the increase in the lattice parameters on doping. The electronic structure of the doped structures shows the difference in the localization of the 4f states in the different structures. This localization defines their optical and electrical properties. Rare earth metal doped ZnO showed a low absorption and a reflectivity less than 7% in the visible range. A shift in blue absorption was observed on doping, revealing an increase in the band gap as compared to pure ZnO. The electrical conductivity has been improved, especially for Ce doped ZnO. This study confronted with experimental results and can be considered doped ZnO are promising for applications in optoelectronics and optical window in solar cells.

**Acknowledgement**

*D. P. Rai acknowledges Core Research Grant from Department of Science and Technology SERB (CRG DST-SERB, New Delhi India) via Sanction no. CRG/2018/000009(Ver-1).*